\input{aipcheck}

\documentclass[
    ,final            
  ]
  {aipproc}

\layoutstyle{6x9}

\newcommand{\ofo}{0509$-$67.5}

\begin{document}

\title{Star Formation Around the Youngest Supernova Remnants in
the Large Magellanic Cloud: Implications for Type Ia Supernova Progenitors}


\author{Carles Badenes}{ address={Department of Astrophysical Sciences, Princeton University, Peyton Hall, Ivy Lane,
    Princeton NJ 08544-1001; \textit{Chandra} Fellow} }

\author{Jason Harris}{ address={National Optical Astronomy Observatory, 950 North Cherry Ave., Tucson, AZ 85719} }

\author{Dennis Zaritsky}{ address={Steward Observatory, 933 North Cherry Ave., Tucson, AZ 85721} }

\author{Jos\'e Luis Prieto}{ address={Department of Astronomy, Ohio State University, McPherson Laboratory, 140 W. 18th
    Avenue. Columbus, OH 43210} }

\begin{abstract}
  We use the star formation history map of the Large Magellanic Cloud recently published by Harris \& Zaritsky to study
  the sites of the youngest Type Ia supernova remnants. We find that most Type Ia remnants are associated with old,
  metal-poor stellar populations, with little or no recent star formation. These include SNR \ofo\, which is known to
  have been originated by an extremely bright SN 1991T-like event, and yet is located very far away from any star
  forming regions. It is very unlikely that this bright Type Ia SN had a young stellar progenitor. The Type Ia remnant
  SNR N103B, however, is associated with vigorous star formation activity in the last 100 Myr, and might have had a
  relatively younger and more massive progenitor.
\end{abstract}

\maketitle


\section{A New Way To Constrain The Properties Of Supernova Progenitors}

After decades of efforts, the fundamental properties of Type Ia supernova (SN) progenitors still remain obscure (see
\cite{maoz08:fraction_intermediate_stars_Ia_progenitors} and references therein). The reason is simple: although
hundreds of new Type Ia SNe are discovered each year, their host galaxies are too far away to either detect the the
faint progenitors in pre-explosion images \cite{maoz08:Ia_progenitor_search_NGC1316} or study the stellar populations
around the explosions with enough detail. Most of the times, all that can be learned from the study of the hosts is an
average age and metallicity for the \textit{entire} galaxy \cite{prieto08:SN_Progenitors_Metallicities}.

Focusing on Local Group galaxies has the advantage that the stellar populations can be resolved and studied in exquisite
detail. Waiting for SNe to explode in these galaxies is obviously not practical, but there is no need for that. Recent
advances in our understanding of the X-ray emission from Type Ia supernova remnants (SNRs)
\cite{badenes03:xray,badenes05:xray} have made it possible to determine the fundamental properties of Type Ia SNe by
studying their SNRs (see \cite{badenes06:tycho} for an example). In some cases it has been possible to verify the
results obtained from the SNR with the spectroscopy of the light echoes associated with the parent SN itself
(e.g. \cite{rest08:0509,badenes08:0509} for the case of SNR \ofo). We take advantage of these techniques to study the
stellar star formation history (SFH) map of the LMC published by \citet{harris08:LMC_SFH}, which was produced by
applying the StarFISH code \cite{harris01:StarFISH} to the data from the Magellanic Clouds Photometric Survey
\cite{zaritsky04:MCPS}.

\begin{figure}
  \includegraphics[height=.3\textheight,angle=90]{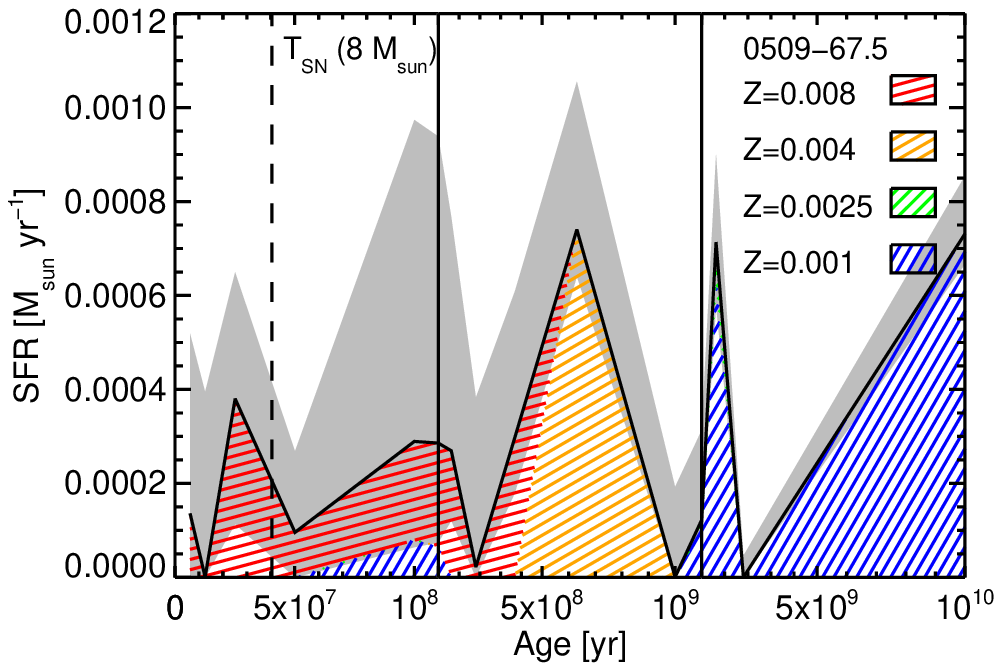}
  \includegraphics[height=.3\textheight,angle=90]{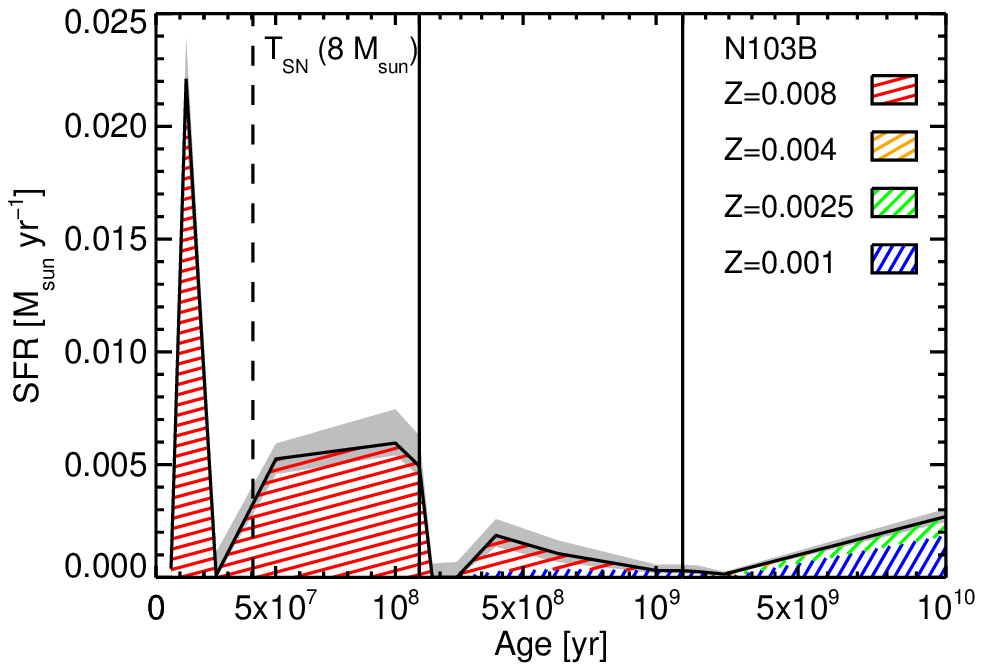}
  \caption{Local SFHs around the Type Ia SNRs \ofo\ (left panel) and N103B (right panel), broken down in four
    metallicity bins. The grey area corresponds to the error bar on the total SFH. \label{fig-1}}
\end{figure}


Our results will be presented in a forthcoming publication (Badenes et al. 2009, in preparation). In Figure \ref{fig-1} we
show the SFHs associated with two of the four Type Ia SNRs in our study. From both its X-ray spectrum
\cite{badenes08:0509} and the SN light echo \cite{rest08:0509}, SNR \ofo\ is known to have been originated by an
exceptionally bright, SN 1991-T like Type Ia SN. However, the stellar population around it is very old (average age 7.9
Gyr) and metal-poor (average $Z$ 0.0014). The SNR is located in a region of the LMC called the Northwestern Void, very
far away from any sites of recent star formation \cite{harris08:LMC_SFH}. This is in contrast to the prevalent view that
bright Type Ia SNe come from younger and possibly more massive progenitors
\citet{gallagher05:chemistry_SFR_SNIa_hosts,scannapieco05:A+B_models}. The situation for SNR N103B is very
different. The SFH is dominated by a vigorous episode of recent star formation, with one extended peak between 50 and
100 Myr ago and a more brief outburst 12 Myr ago. This Type Ia SNR might have had a more massive and younger progenitor
than SNR \ofo.




\bibliographystyle{aipproc}   


\IfFileExists{\jobname.bbl}{}
 {\typeout{}
  \typeout{******************************************}
  \typeout{** Please run "bibtex \jobname" to optain}
  \typeout{** the bibliography and then re-run LaTeX}
  \typeout{** twice to fix the references!}
  \typeout{******************************************}
  \typeout{}
 }

\end{document}